\documentclass[fleqn,10pt]{wlscirep}

\usepackage{lineno}
\usepackage[english]{babel}
\usepackage[utf8]{inputenc} 
\usepackage[T1]{fontenc} 
\usepackage[none]{hyphenat}
\usepackage{multirow}
\usepackage{caption}
\usepackage{subcaption}
\usepackage{textcomp}
\usepackage[square,sort,comma,numbers]{natbib}
\usepackage{amsmath}
\usepackage{graphicx}
\usepackage[colorinlistoftodos]{todonotes}
\usepackage[colorlinks=true, allcolors=blue]{hyperref}
\usepackage{xcolor}
\usepackage{amssymb}
\usepackage[capitalise, nameinlink, noabbrev]{cleveref}
\usepackage{xspace}
\usepackage{color,soul}
\usepackage{enumitem}
\newlist{inlinelist}{enumerate*}{1}
\setlist*[inlinelist,1]{%
  label=(\roman*),
}
\usepackage{tcolorbox}
\usepackage{pdfpages}

\definecolor{pink}{rgb}{1,0.33,0.5}

\newcommand{\DatasetName}{WEMAC\xspace}

\newcommand{\bindi}{Bindi\xspace}

\nolinenumbers

\title{WEMAC: Women and Emotion Multi-modal Affective Computing dataset}

\author[1,5,$\dag$,*]{Jose A. Miranda Calero}
\author[2,$\dag$]{Laura Gutiérrez-Martín}
\author[3, 5, $\dag$,*]{Esther Rituerto-Gonzalez}
\author[2, 5]{Elena Romero-Perales} 
\author[4]{Jose M. Lanza-Gutiérrez} 
\author[3, 5]{Carmen Peláez-Moreno} 
\author[2, 5]{Celia López-Ongil} 
\affil[1]{Embedded Systems Laboratory, Ecole Polytechnique Fédérale de Lausanne (EPFL), Switzerland}
\affil[2]{Department of Electronics, University Carlos III of Madrid (UC3M), Spain}
\affil[3]{Department of Signal Theory and Communications, University Carlos III of Madrid (UC3M), Spain}
\affil[4]{
Department of Computer Science, Universidad de Alcal\'a (UAH), Spain
}
\affil[5]{Gender Studies Institute, University Carlos III of Madrid (UC3M), Spain}

\affil[*]{corresponding author(s): Jose A. Miranda (jose.mirandacalero@epfl.ch), Esther Rituerto-Gonzalez (erituert@ing.uc3m.es)}

\affil[$\dag$]{these authors contributed equally to this work}

\begin{abstract}

WEMAC is a unique open multi-modal dataset that comprises physiological, speech, and self-reported emotional data records of $100$ women, targeting Gender-based Violence detection. Realistic emotions were elicited through visualizing a validated video set using an immersive virtual reality headset. The physiological signals captured during the experiment include blood volume pulse, galvanic skin response, and skin temperature. The speech was acquired right after the stimuli visualization to capture the final traces of the perceived emotion. Subjects were asked to annotate among $12$ categorical emotions, several dimensional emotions with a modified version of the Self-Assessment Manikin, and liking and familiarity labels. The technical validation proves that all the targeted categorical emotions show a strong statistically significant positive correlation with their corresponding reported ones. That means that the videos elicit the desired emotions in the users in most cases. Specifically, a negative correlation is found when comparing \textit{fear} and \textit{not-fear} emotions, indicating that this is a well-portrayed emotional dimension, a specific, though not exclusive, purpose of WEMAC towards detecting gender violence.

\end{abstract}
\begin{document}




\maketitle



\section*{Background \& Summary}

Gender-based Violence (GBV) is a violation of human rights and fundamental liberties as declared by the United Nations in $1993$ \cite{unitednations1993}. This declaration states that GBV is any act of physical, sexual, or psychological violence directed toward the female gender. A datum that helps ponder its impact on society shows that more than $27\%$ of ever-partnered women aged between $15$ and $49$ experienced physical or sexual violence by intimate partners from $2000$ to $2018$ \cite{global_gbv}. Another worrying statistic is that, in $2020$, approximately $47,000$ women and girls were killed worldwide by their intimate partners or other family members, meaning a woman or girl is killed by someone in her own family every $11$ minute \cite{UNODC_2021}. On this basis, it can be understood that GBV is an urgent problem worldwide that should be addressed by science and society. From a sociological point of view, particular emphasis on education is essential to eradicate, combat, and prevent GBV, but it requires several generations to produce a change in society. In the meantime, technology can be fundamental for helping prevent and combat GBV \cite{genderviolencetechnology} and empowering women. With this focus, the multidisciplinary team UC3M4Safety\footnote{https://www.uc3m.es/institute-gender-studies/UC3M4Safety} was created in $2017$ to propose Bindi, an inconspicuous autonomous system powered by artificial intelligence and deployed under the Internet of Things (IoT) paradigm. The goal of Bindi is to automatically report when a woman is in a GBV-related risk situation to trigger a protection protocol \cite{miranda2022bindi}. This risk situation identification is performed by detecting fear-related emotions in the user through a multi-modal intelligent engine feeding with physiological and audio data captured by a pair of wearable devices: a bracelet and a pendant.

One of the main shortcomings of training this specific \textit{fear} detection system for \bindi is the lack of adequate datasets.  
Over the last decades, several datasets were published providing emotional labels together with auditory and physiological variables, such as DEAP \cite{DEAP}, MAHNOB \cite{MAHNOB}, WESAD \cite{WESAD}, AMIGOS \cite{AMIGOS}, {FAU, Reg, and Ulm TSST Corpora} \cite{alice_tsst}, and BioSpeech \cite{biospeech}. However, the target of these datasets is the classification of a generic set of emotions instead of focusing on \textit{fear} detection. This strategy makes it hard to obtain a robust model due to the lack of \textit{fear} samples. Moreover, gender perspective was not considered either, in spite of stimuli interpretation being strongly affected by gender \cite{genderemotions}. Instead, a sufficient number of women volunteers and a balanced \textit{fear} and \textit{non-fear} set of emotions should be considered to fit the target users of the GBV application. 

To fill this gap, UC3M4Safety generated the UC3M4Safety Database \cite{blanco2021uc3m4safety}, which includes the Women and Emotion Multi-modal Affective Computing (WEMAC) and audiovisual stimuli datasets, as listed in Table \ref{tab:database_hierarchy}. This paper presents WEMAC, a collection of physiological and audio data captured in a virtual reality set-up where women volunteers are exposed to a subset of audiovisual stimuli previously rated by experts judges and an online crowd-sourcing procedure \cite{db_estimuli_video} \cite{db_estimuli_annotations}. For each volunteer, WEMAC includes the answers to an initial questionnaire \cite{wemac_questionnaire}, physiological signals \cite{wemac_physiological}, features extracted from the volunteer's speech data to ensure confidentiality \cite{wemac_audiofeatures}, and self-reported emotional labels \cite{wemac_labels}.

\begin{table}[h]
\centering
\def\arraystretch{1.5}
\begin{tabular}{c|c|c|c}
\textbf{Database}& \textbf{Datasets}& \textbf{Conditions}& \textbf{Participants}
\\ 
\hline
\multirow{6}{*}{\begin{tabular}[c]{@{}c@{}}UC3M4Safety \\ Database  \cite{blanco2021uc3m4safety}
\end{tabular}} & 
Audiovisual  Stimuli: Videos \cite{db_estimuli_video}& \multirow{2}{*}{Online crowd-sourcing}  
& General public and  
\\
& Audiovisual  Stimuli: Emotional Ratings  \cite{db_estimuli_annotations}&     &                    expert judges\\ 
\cline{2-4}& 
\DatasetName: Biopsychosocial Questionnaire \cite{wemac_questionnaire} & \multirow{4}{*}{\begin{tabular}[c]{@{}c@{}}Laboratory \end{tabular}} & \multirow{4}{*}{Women volunteers} 
\\
& \DatasetName: Physiological Signals \cite{wemac_physiological} &     &             \\
&\DatasetName: Audio Features \cite{wemac_audiofeatures} &     &                     \\&  \DatasetName: Self-reported Emotional Annotations \cite{wemac_labels}  &     &                     \\ 
\hline
\end{tabular}
\caption{Hierarchy and subdivisions of the UC3M4Safety Database datasets.}
\label{tab:database_hierarchy}
\end{table}

WEMAC offers several advantages over other state-of-the-art datasets: 
\begin{inlinelist}[label = \arabic*)]
    \item the use of immersive technology (virtual reality) to elicit emotions, which offers a high degree of correlation between the research conditions and the emotional phenomenon under study;
    \item a high number of volunteers ($100$), higher than any other public dataset of this type to the best of our knowledge; 
    \item a specifically designed modification of the labeling methodology to consider gender perspective by changing the design of the original Self-Assessment Manikins (SAMs) \cite{samClara};
    \item the implementation of an online recovery process to ensure a physiological stabilization between stimuli, and 
    \item the usage of different sensory systems to provide a heterogeneous set-up approach.
\end{inlinelist}

WEMAC is intended, but not limited, to address research questions related to
\begin{inlinelist}[label = \arabic*)]
\item affective computing using multi-modal information,
\item the design of solutions to the very challenging problem of GBV, 
\item the understanding of subjective and self-reported emotional labeling, and
\item \textit{fear} classification in women.

\end{inlinelist}


\section*{Methods}

\subsection*{Ethics statement}
\label{sec:ethics}

The experimentation was approved by and performed following the guidelines and regulations of the Ethics in Research Committee of University Carlos III Madrid. The approval was granted considering the circumstances of the research project entitled \textit{Integral protection of gender-based violence victims using multimodal affective computing}\footnote{The original Spanish title is \textit{Protección Integral de las víctimas de violencia de género mediante Computación Afectiva multimodal}
(EMPATIA-CM) with reference Y2018/TCS-5046 and funded by the programme \textit{Proyectos
Sinérgicos I+D (Comunidad de Madrid, Consejería de Ciencia, Universidades e Innovación)}, whose principal investigator is Dr. Celia López-Ongil.}.

The submission to the Ethical Committee covered essential topics for the development of the experiments. Among others, the adequacy of the volunteers' informed consent, the research goals and plans, the data management and de-identification procedures, and the compliance with the European General Data Protection Regulation (GDPR). 
The aforementioned written informed consent asserts that the volunteers 
\begin{inlinelist}[label = \arabic*)]
\item were aware of the research, its objectives, purposes, and how their data was planned to be used;
\item were informed regarding the experimental procedure, the possibility of refusing to participate in the research at any point, and the right to request data erasure;
\item could ask any question during the experiments. 
\end{inlinelist}
Moreover, they granted permission for processing their personal data to the extent necessary for the implementation of the research project, including sharing with other researchers their physiological recording and speech features, as well as the initial questionnaires and self-reported annotations.

\subsection*{Participants}
\label{sec:participants}

The volunteers were recruited by using different communication channels, including social networks (such as Facebook, Twitter, or Instagram), Women's Associations, and the collaboration of the municipality of Leganés and Getafe (Madrid, Spain), resulting in an initial selection of $144$ Spanish-speaking women volunteers. This initial set was reduced due to technical problems during the recording or sickness caused by the virtual reality set-up involving the interruption of the experiment. The final set includes $100$ women volunteers aged between $20$ and $77$ (with an average of $39.92$ and a standard deviation of $14.26$). For the purpose of covering a wide age range, the recruitment process requested a balanced number of volunteers in five age groups defined as G1 ($18-24$) for which we recruited $22$ volunteers, G2 ($25-34$) with $18$ volunteers and G3 ($35-44$), G4 ($45-54$), and G5 ($\geq 55$) that include $20$ volunteers each. 

\subsection*{Stimuli}
\label{sec:stimuli}

The selection of audiovisual clips to record WEMAC is the result of a thorough study done with the purpose of collecting a high-quality set of audiovisual stimuli able to trigger realistic emotions under a controlled scenario \cite{preetiquetado}. 
To this end, $370$ samples with emotional content were initially collected on the Internet from commercial films, series, documentaries, short films, commercials, and video clips. The set was labeled with the advice of a panel of experts seeking to elicit the following $12$ discrete emotions: \textit{joy, sadness, surprise, contempt, hope, fear, attraction, disgust, tenderness, anger, calm}, and \textit{tedium}. After that, the team discarded those clips over two minutes long, those needing context to be understood, or that elicited more than one target emotion. Afterward, the resulting $162$ clips were surveyed in an online crowd-sourcing poll to be labeled with discrete emotion categories after its visualization, getting $1,520$ volunteer annotators ($929$ women and $591$ men). A further selection was performed by considering two conditions: 
\begin{inlinelist}[label = \arabic*)]
\item at least $50\%$ of the volunteers (considering both genders or at least $50\%$ of one gender individually) labeled the clip with the same discrete emotion;
\item the rest of the emotions labeled only reached a maximum of $30\%$ agreement to avoid confusion with the target emotion
\end{inlinelist}. After this selection, none of the videos previously selected for \textit{attraction, contempt, hope,} and \textit{tedium} had greater than $50\%$ agreement among the responses obtained, so the final set of videos covers a list of 8 target emotions \textit{joy, sadness, surprise, fear, attraction, disgust, tenderness, anger,} and \textit{calm}. Finally, some videos were discarded to obtain an even distribution between \textit{fear} and \textit{non-fear} emotions. It resulted in the Audiovisual Stimuli dataset \cite{db_estimuli_video} with $42$ clips and a distribution of $44.44\%$ for \textit{fear} and $55.55\%$ for \textit{no-fear} \cite{db_estimuli_annotations}.

Applying the whole Audiovisual Stimuli dataset in WEMAC  would be unfeasible due to the excessive duration of the resulting experimentation. Thus, two batches of videos are generated so that the experimental procedure lasts $1$ to $1.5$ hours per volunteer. The selection criterion for the videos in these two batches is based on three premises: the emotional highest discrete labeling agreement, targeting for an adequate laboratory experiment duration, and a balanced distribution of \textit{fear} and \textit{non-fear} categories within the four quadrants in the Pleasure-Arousal space \cite{fontaine}. Table \ref{tableBatch} shows details about the $14$ selected videos for each batch in WEMAC, including the stimuli identification (Stimuli ID, with the same notation as followed in the Audiovisual Stimuli dataset \cite{db_estimuli_video}), the visualization order in the experimentation, the emotion label reported in the online crowd-sourcing study, the video duration, and the visualization format.

\begin{table}[h]
\centering
\begin{tabular}{ccccccc}
\textbf{Stimuli ID} & \textbf{Visualization order} & \textbf{Emotion label} &  \textbf{Duration} & \textbf{Format} & 
\textbf{Batch} \\
V01 & 1 & Joy  & 1'26'' & 2D & 1 \\
V15 & 2 & Fear & 1'20" & 3D & 1 \\
V36 & 3 & Sadness & 1'59" & 2D & 1 \\
V08 & 4 & Anger & 1'03" & 3D & 1 \\
V28 & 5 & Fear & 1'35" & 2D & 1 \\
V40 & 6 & Calm & 1' & 3D & 1 \\
V09 & 7 & Anger & 1' & 2D & 1 \\
V19 & 8 & Fear & 23" & 2D & 1 \\
V52 & 9 & Disgust & 40" & 2D & 1 \\
V16 & 10 & Fear & 2' & 3D & 1 \\
V02 & 11 & Joy & 1'41'' & 2D & 1 \\
V27 & 12 & Fear & 1'20" & 2D & 1 \\
V37 & 13 & Gratitude & 1'40" & 2D & 1 \\
V24 & 14 & Fear & 1'27" & 2D & 1 \\
V22 & 1 & Fear & 1'52" & 2D & 2 \\
V04 & 2 & Joy  & 1'28'' & 2D & 2 \\
V11 & 3 & Fear  & 46'' & 2D & 2 \\
V34 & 4 & Sadness & 45" & 2D & 2 \\
V13 & 5 & Fear & 1'33'' & 3D & 2 \\
V41 & 6 & Calm & 1' & 2D & 2 \\
V10 & 7 & Anger & 1'59'' & 2D & 2 \\
V25 & 8 & Fear & 1'14" & 2D & 2 \\
V33 & 9 & Disgust & 1'36" & 2D & 2 \\
V14 & 10 & Fear & 2' & 3D & 2 \\
V07 & 11 & Surprise & 1'41'' & 2D & 2 \\
V26 & 12 & Fear & 1'06" & 2D & 2 \\
V77 & 13 & Gratitude & 1'30" & 2D & 2 \\
V12 & 14 & Fear & 1'59'' & 3D & 2
\end{tabular}
\caption{List of selected audio-visual stimuli used within the \DatasetName\ Dataset.}
\label{tableBatch} 
\end{table}

\subsection*{Measures}
\label{sec:measures}

Volunteers' annotations are collected in two instants: prior to the experiment and during the experimentation. Before the experiment, each volunteer is provided with informed consent, a personal data form, and a general questionnaire to supply additional information related to cognition, appraisal, attention, personality traits, gender, and age. In this regard, the general questionnaire collects age group, recent physical activity or medication that can alter the physiological response of the participant, self-identified emotional burdens due to work, economic and personal situation, and mood biases (fears, phobias, and previous traumatic experiences).

During the experimental protocol, in addition to physiological information, the following self-assessment annotations are obtained after each visualized stimulus:
\begin{itemize}
    \item Speech-based labeling: two questions regarding each video stimulus are asked immediately after its visualization. Their goal is to make the user relive the emotions felt during viewing. Some examples are "What did you feel during the visualization of the video?" or "Could you describe what happened in the video using your own words?". The answer is stored as an audio signal. Note that both questions and answers are in Spanish.
    \item Valence, Arousal, and Dominance: annotated using a $9$-point Likert scale supported by modified SAMs. 
    Details about the redesign process of the SAMs can be found in \cite{samClara}.
    \item Familiarity with the emotion felt, the situation displayed in the clip, and the specific clip: annotated in three different questions. The two first consider a $9$-point Likert scale, whereas the last one considers a binary yes-no option.
    \item Liking of the video: annotated through a binary yes-no question.
    \item A discrete emotion out of a total of $12$ (\textit{joy, sadness, surprise, contempt, hope, fear, attraction, disgust, tenderness, anger, calm,} and \textit{tedium}) \cite{preetiquetado}. 
\end{itemize}

\subsection*{Apparatus}
\label{sec:apparatus}
The equipment employed to capture the physiological and speech data is as follows:
\begin{itemize}  
    \item The BioSignalPlux\footnote{https://biosignalsplux.com/products/kits/researcher.html} research toolkit system. It is a commonly used device to acquire different physiological signals in the literature \cite{liu2021csl,liu2022practical,carvalho2022heartbeat,harjani2019analysis}. More specifically, we capture finger Blood Volume Pulse (BVP), ventral wrist Galvanic Skin Response (GSR), forearm Skin Temperature (SKT), trapezoidal Electromyography (EMG), chest respiration (RESP), and inertial wrist movement through an accelerometer. 
    \item The Oculus Rift\textregistered\ S VR Headset\footnote{https://www.oculus.com/rift-s/}. Its embedded microphone captures the speech signal produced during the speech-based labeling at a sampling rate of $48~kHz$ mono and a depth of $16$ bits. This device also guides volunteers through the study. To this end, an interactive virtual reality environment developed in Unity software\footnote{https://unity.com/} presents stimuli and collects self-assessments. More details about this virtual environment can be found in \cite{tfg_laura}.
    \item Additionally, two in-house sensory systems are employed. On the one hand, the \bindi's bracelet \cite{iot_bbddlab} measures dorsal wrist BVP, ventral wrist GSR, and forearm SKT. The hardware and software particularities of this device are detailed in \cite{cpsDTE,dcisBVP,dcisGSR}. The previously mentioned BioSignalPlux toolkit is employed as a golden standard to analyze the performance of its sensors due to its experimental nature. BVP and GSR signals from BioSignalPlux and \bindi\ were successfully compared and correlated with \bindi\ in \cite{dcisBVP} and \cite{dcisGSR}. On the other hand, a GSR sensor to be integrated into the next version of the \bindi\ bracelet is used. Its hardware and software particularities are detailed in \cite{newGSRbindi}.
\end{itemize}
Note that the synchronization of all the different sensors acquisition together with the stages of the experiment is performed using a laptop (MSI GE75 Raider 8SE-034ES) running a Unity\textregistered\ framework-based program. On the one hand, the BiosignalPlux device connection is configured using the OpenSignals (r)evolution software\footnote{https://support.pluxbiosignals.com/article-categories/opensignals/}, and its TCP/IP module is used to facilitate the data exchange between this platform and the Unity\textregistered\ framework. On the other hand, the additional in-house sensory systems are wirelessly connected to the laptop using Bluetooth Low Energy dongles. The information storage is divided by scenes and marked individually with a timestamp set by the environment employed. The sampling frequency for all physiological signals is $200~Hz$.

\subsection*{Procedure}

\begin{figure}[!htbp]
   \centerline{
   \includegraphics[scale=0.45]{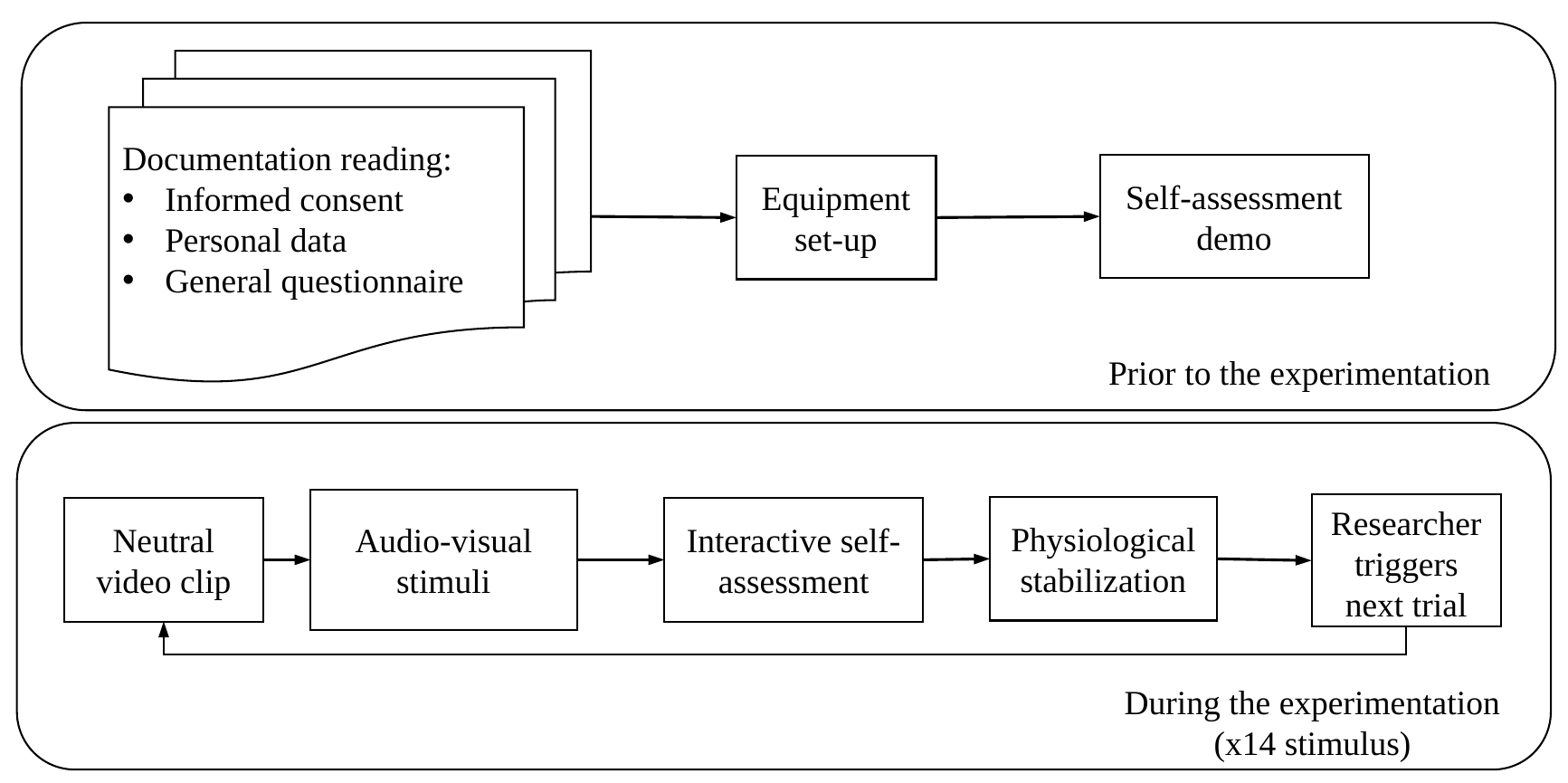}}
    \caption{Experimental methodology followed during the development of the \DatasetName\ dataset. Prior to and during the experimentation for each volunteer.}
    \label{methodologyUc3m4safety}
\end{figure}

{The study was conducted between October $2020$ and April $2021$.} It took place in the
Electronics Technology Department at the School of Engineering of the Universidad Carlos III de Madrid, Spain. The experimental methodology designed to be applied for each volunteer is schematized in Figure~\ref{methodologyUc3m4safety}. During this experiment, at least one researcher and one psychologist remained in the room at the disposal of the participants in case they needed any help.

Upon arrival, participants were informed about the experimental procedure. Then, they signed the informed consent, filled out the personal data form, and answered the general questionnaire. Next, participants listened to instructions regarding the experiment.
The participants were requested to avoid unnecessary actions or movements during the experiment (e.g., turning the wrist). They were also informed that they could skip any clip or quit the experiment at any
time. Once the procedure was clear to the participants, the sensors were set up, as well as the Virtual Reality Headset.  
Next, the participants followed a tutorial to get used to the headset, joystick, interactive screens, and the particularities of the different annotations.
The main part of the experiment consisted of fourteen iterations (one per stimulus) of 
\begin{inlinelist}[label = \roman*)]
\item a neutral clip sampled from  \cite{rottenberg2007emotion}, 
\item the visualization of the emotional film clip, 
\item different interactive screens of self-assessment annotations, and 
\item 3D recovery landscape scenes. 
\end{inlinelist}

The full experiment, including documentation reading, equipment set-up, tutorial, and visualization of the clips, lasted around $80$ minutes per participant, depending on the time spent on the questionnaires.

\subsection*{Data processing and cleaning}
\label{sec:data_proc}
The whole set of physiological signals captured for the $100$ volunteers has a total duration of 34:51:50 (HH:MM:SS), whereas the audio signals altogether last 12:19:25 since they are only saved during the speech-based annotation.

\subsubsection*{Physiological Signals}
The signals being released are the ones acquired by the BioSignalPlux research toolkit. Specifically, the raw and filtered BVP, GSR, and SKT signals captured during every video visualization are provided. The preprocessing is as follows: 
\begin{enumerate}
    \item Two main filters are applied to the BVP signal due to the noise problems observed. On the one hand, high-frequency noise is filtered out using a direct-form low-pass Finite Impulse Response (FIR) filter with $6~dB$ at $3.5~Hz$. Note that a Hamming window is used during the filter design process to minimize the first side lobe properly. On the other hand, the residual baseline-wander or low-frequency drift effect presented in the signal is removed using a forward-backwards low-pass Butterworth Infinite Impulse Response (IIR) filtering stage. Specifically, the forward-backwards technique handles the non-linear phase of such filters.
    \item For the GSR and SKT signals, a basic FIR filtering with $2~Hz$ cut-off frequency is applied. After that, this filtered output is downsampled to $10~Hz$ and also processed with both a moving average and a moving median filter. The former used a $1~s$ long window and helped to reduce the high noise residual after the initial FIR, whereas the latter employed a $0.5~s$ window and dealt with the rapid transients.
\end{enumerate}

\subsubsection*{Speech Signals}

The speech signals recorded have a duration ranging from $20~s$ to $60~s$ and mostly contain speech. 
Since the release of the raw speech signals is not possible  due to ethics and privacy issues \footnote{Regulation (EU) 2016/679 of the European Parliament and of the Council of 27 April 2016 on the protection of natural persons with regard to the processing of personal data and on the free movement of such data, and repealing Directive 95/46/EC (\textit{General Data Protection Regulation})}, we preprocessed the speech signals and extracted both low and high-level features so that the research community can analyze and work with them.

The preprocessing is as follows. A low pass filter at $8~kHz$ is applied, where most of the energy is concentrated. A high pass filter with a cut-off frequency of $50~Hz$ is also applied to remove the electrical power line interference. Afterward, the audio amplitude is normalized per participant to fit the range $[-1, 1]$. That means the normalization per participant was performed using all her audio signals. 
Next, downsampling at $16~kHz$ is performed with the \textit{librosa} Python library \cite{librosa} to facilitate the handling of the signals. Finally, the signals are padded with zeros to fill in the incomplete last second.

Different Python toolkits are used to extract information from the preprocessed signal at a window size of $1~s$ without overlapping. 
We follow a similar approach to the one followed in the MuSe Challenge $2021$ \cite{muse_challenge} for the feature and embedding extraction of the audio signals. That is:
\begin{enumerate}
\itemsep0em
\item \textit{librosa} \cite{librosa} \cite{librosa_software}: 
$19$ features are extracted by the librosa Python toolkit ($13$ Mel-Frequency Cepstral Coefficients, Root Mean Square or Energy, Zero Crossing Rate, Spectral Centroid, Spectral Roll-off, Spectral Flatness, and Pitch) at a window size of $20~ms$ and a hop size of $10~ms$. Also, the mean and standard deviation for each of the $19$ features are computed for every $1~s$ window. As a result, $38$ speech features are computed.

\item \textit{eGeMAPS} \cite{egemaps}: $88$ Low-Level Descriptors (LLDs) related to speech and audio are extracted through the \texttt{openSMILE} Python toolkit \cite{opensmile} on its default configuration, i.e., F0, harmonic features, HNR, jitter, and shimmer are computed with a $60~ms$ window. Loudness, spectral slope, formants, harmonics, Hammarberg Index, and Alpha ratio are computed with a $25~ms$ window. They are all further averaged in $1~s$ windows.

\item \textit{ComParE}: $6,373$ features used in the \text{ComParE} $2016$ challenge \cite{compare} are extracted by using the \texttt{openSMILE} Python toolkit. The default window and step sizes are used as in \cite{florian2015thesis}, i.e., F0, jitter, and shimmer are computed with a window size of $60~ms$ and a step of $10~ms$. All other LLDs are computed with a window size of $20~ms$ and a step of $10~ms$. Then all features are further averaged in $1~s$ windows.

\item \textit{DeepSpectrum} \cite{amiriparian2017Snore}: This toolkit is used for the extraction of audio embeddings based on different deep neural network architectures trained with ImageNet \cite{imagenet}. Specifically, two different configurations are considered, and therefore two embeddings sets are extracted, i.e., \text{ResNet50} and the output of its last Average Pooling layer (\textit{avg\_pool}), resulting in $2048$-dimensional embeddings, and \text{VGG-19} and its last Fully Connected layer (\textit{fc2}), resulting in $4096$-dimensional embeddings.

\item \textit{VGGish}: The $128$-dimensional embeddings from the output layer of the \text{VGG-19} network trained for AudioSet \cite{audioset} are also included.

\item \textit{PASE+}: The $256$-dimensional features from the \text{PASE+} (\textit{Problem Agnostic Speech Encoder+}) \cite{paseplus} encoder network, used as speech feature extractor, are also provided.
\end{enumerate}


\section*{Data Records}
The collected data in WEMAC (physiological signals, audio features, self-reported annotations, and general questionnaire answers) are available in e-cienciaDatos portal \cite{blanco2021uc3m4safety}, a research data repository held by Madroño Consortium (composed of Universities in the Madrid-Spain region and member of the Harvard Dataverse Network, accessible for major publishers). 
The information is split into four sections according to their typology, where each set contains a `README.txt' file that explains each specific content. 

\begin{enumerate}
\itemsep0em
\item \textit{Bio-psycho-social questionnaire and informed consent} \cite{wemac_questionnaire}: a tabular data CSV (Comma-Separated Values) file with $100$ rows corresponding to each of the volunteers and $14$ columns corresponding to the information completed in the biopsychosocial questionnaire at the beginning of the experiment. This information includes: (1) volunteer identifier; (2) the age group in which she was included; (3) if she had signed the informed consent; (4) the number of children she had if any; (5) if she had had energy drinks before the experiment; (6) if she was taking medicines; (7) if she usually did physical exercise; (8) if she considered herself fearful; (9) if she had any deepest fear; (10) if she was in a stressful period; (11) if she had experienced any traumatic experiences; (12) if social interaction made her feel anxious; (13) if she considered having a tendency to be worried; (14) if she had ever felt fear of aggression; and (15) if she had ever felt a threat to her sexual integrity. 

\item \textit{Self-reported annotations} \cite{wemac_labels}: They contain the emotional labeling reported by the participants after watching each of the 
$14$ videos in the experiment. 
The data are stored in one CSV file, that contains $14$ columns and $1,400$ rows ($100$ volunteers $\times$ $14$ clips). Regarding the columns, $5$ of them refer to information about the volunteer (identifier and age group) and the video (batch, display position, and video code number), whereas $9$ of them refers to the self-assessment provided (arousal, valence, dominance, liking, reported and target discrete emotions, and familiarity scores for the emotion, the situation, and the clip).

\item \textit{Physiological signals} \cite{wemac_physiological}: BVP, GSR, and SKT physiological signals captured during the experimentation by the BioSignalPlux research toolkit are provided in a binary MATLAB® file (.mat). 
It contains a cell array with $100$ rows (one per volunteer) and $14$ columns (one per video). Each cell contains four fields: volunteer identifier, clip or trial identifier, filtering indicator, and an inner cell array (with the physiological data associated with that specific clip and volunteer). 

\item \textit{Speech features} \cite{wemac_audiofeatures}: The speech features contain $6$ folders: \texttt{librosa},  \texttt{eGeMAPS},  \texttt{Compare}, 
 \texttt{Deepspectrum-Resnet50},  \texttt{Deepspectrum-VGG19},  and  \texttt{VGGish}. They correspond to the six feature sets described in the Data Processing and Cleaning section for Speech Signals. Each folder includes a CSV file per audiovisual stimuli and volunteer. Each CSV has as many columns as the number of features calculated, where an additional first column is also included referring to the timestamp (in seconds). The number of rows fits with the number of seconds the speech signal lasts.


\end{enumerate}

Besides data collection, additional informative documents regarding, for instance, sensor placement during the experiment or the self-assessment instructions, are included in each folder of the data repository.


\section*{Technical Validation}
\begin{figure*}[h!]
\centerline{\includegraphics[scale=0.50]{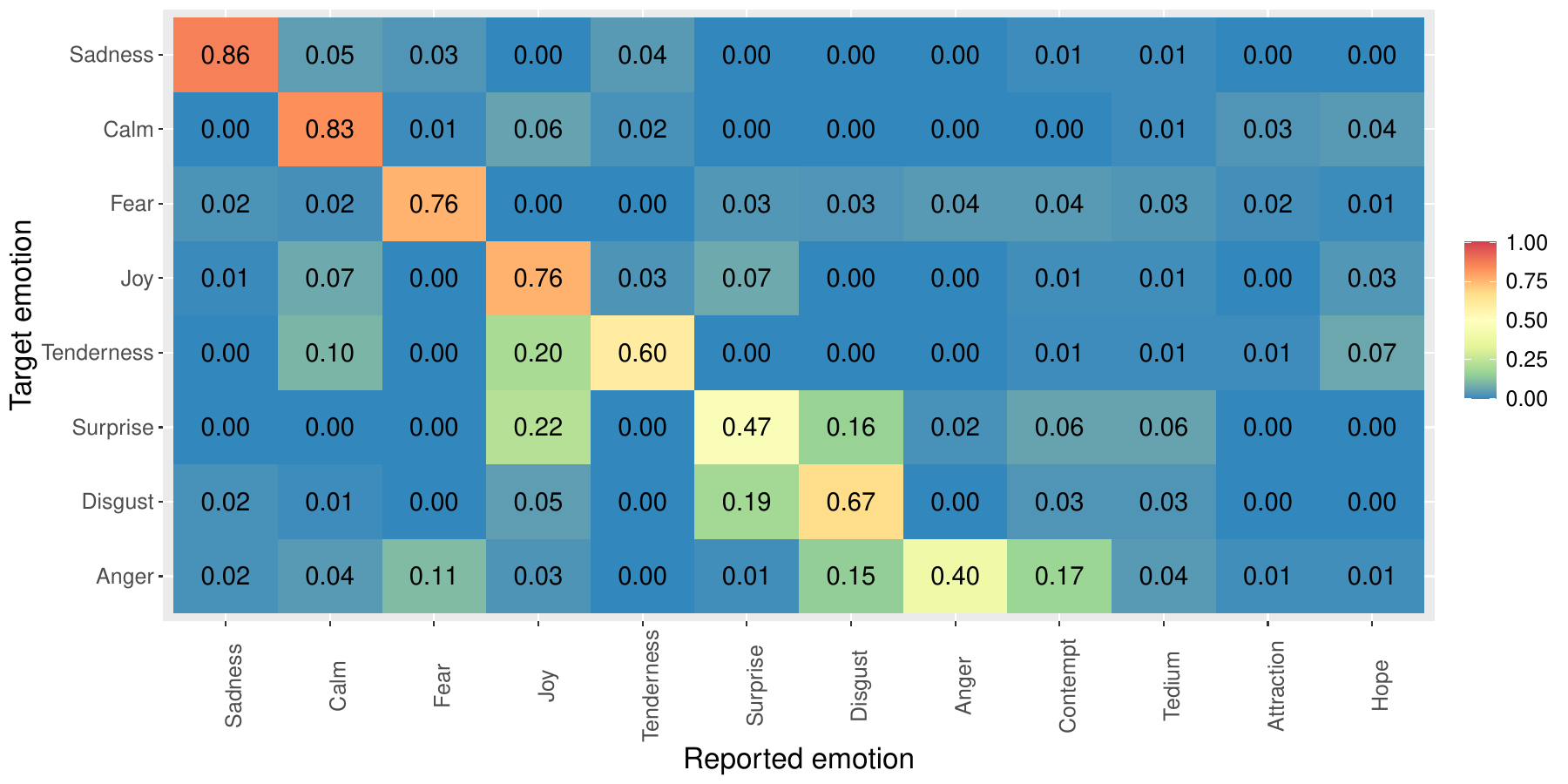}}
\caption{Self-reported emotion labeling distribution (0.00-1.00) comparing the target discrete emotions to the reported ones. It includes $8$ target emotions versus $12$ reported ones.}
\label{figemotiondist}
\end{figure*}

Emotional elicitation and labeling is a complex task, and sometimes the expected (or targeted) emotions are not the ones the volunteers experienced (or reported). The agreement between the target class and the self-reported discrete emotion annotations by the volunteers in this experiment is shown 
in the matrix in Figure~\ref{figemotiondist}, where it is observed as the ratio of times a targeted emotion is identified and felt as such by the volunteers. Thus, a value of 1.00 means a perfect agreement between the targeted emotion and the emotion felt, and 0.00 means no agreement. 
As introduced before, only $8$ of the $12$ emotions initially selected were included in WEMAC (see the Stimuli Section),  although the $12$ emotions were considered for the discrete emotion labeling (see the Measures Section). It means that the number of targeted emotions is smaller than the reported ones in this matrix. Analyzing this figure it can be found that the non-included emotions (\textit{attraction, contempt, hope} and \textit{tedium}) are very scarcely selected with the exception of the $17\%$ of times a stimulus expected to represent \textit{anger} is taken as \textit{contempt}. It is also observed that \textit{sadness, calm, joy} and \textit{fear} are the emotions best identified, being the agreement in the fear emotion especially relevant for the use case. \textit{Tenderness} and \textit{disgust} are also quite well portrayed by the stimuli while \textit{anger} is often taken as \textit{disgust} or \textit{contempt}, and \textit{amusement} as \textit{joy} or \textit{disgust}.

To statistically confirm the associations observed between the target and self-reported discrete emotions, a Pearson's chi-squared test is used as implemented in the \textit{chisq.test()} function in the R software. In this test, the null hypothesis ($H_{0}$) states that the variables (i.e., the target and reported emotions) are independent, meaning there is no relationship between categorical variables. As a result of this test, a p-value equaling $0.0005$ is obtained, resulting in the dependency between categories is confirmed, with a $0.01$ significance level. Then, we perform a study regarding the relationships between the different target and reported individual emotions. To this end, a posthoc analysis with Bonferroni correction for multiple comparisons is performed for each combination of target and reported emotion. This analysis is performed by means of the Pearson standardized residuals (Z-factor) implemented in the \textit{stdres} component of the result of the previous \textit{chisq.test()} function in the R software. The Z-factors appear in Table \ref{tab:standardizedResiduals}, where a strong dependence occurs when the absolute value of the Z-factor is higher than the reference Z-factor ($3.34$).

In such a case, a positive Z-factor means a strong direct dependence and a negative one means a strong inverse dependence.

\begin{table}[h!]
   \centering
   \resizebox{17cm}{!}{
   \begin{tabular}{c c c c c c c c c c c c c}
   \toprule
   \textbf{Targeted}& \multicolumn{11}{c}{\textbf{Reported emotion}} \\
     \cline{2-13}
	   \textbf{emotion}& \textbf{Amusement}& \textbf{Anger}& \textbf{Attraction}& \textbf{Calm}& \textbf{Contempt}& \textbf{Disgust}& \textbf{Fear}& 
	   \textbf{Hope}& \textbf{Joy}& \textbf{Sadness}& \textbf{Tedium}& \textbf{Tenderness} \\ 
    \midrule
    
	\textbf{Amusement}& \textbf{12.95}& -1.23& -0.77& -2.25& 0.59& 2.05& \textbf{-5.43}& -0.92& 2.44&
	-2.06& 1.49& -1.63\\
	\textbf{Anger}& -2.38& \textbf{18.06}& 0.22& -2.30& \textbf{8.08}& \textbf{3.40}& \textbf{-6.87}& -0.33& \textbf{-3.60}& 
	-2.79& 1.00& -2.98\\
    \textbf{Calm}& -2.50& -2.67& 1.81& \textbf{26.71}& -2.23& -3.13& \textbf{-7.69}& 1.92& -1.79&
    -3.00& -1.09& -1.42\\
    \textbf{Disgust}& \textbf{6.14}& -2.67& -1.12& -2.91& -0.72& \textbf{21.99}& \textbf{-7.90}& -1.34& -2.11& 
    -2.22& 0.18& -2.38\\
    \textbf{Fear}& -3.21& -3.18& 1.30& \textbf{-8.93}& -0.71& \textbf{-6.82}& \textbf{28.32}& -2.39& \textbf{-12.54}& \textbf{-6.84}& 1.12& \textbf{-7.98}\\
    \textbf{Joy}& 1.02& \textbf{-3.35}& -1.40& -0.81& -1.96& \textbf{3.93}& \textbf{-9.91}& 1.70*& \textbf{26.37}& 
    -3.34& -1.11& -1.40\\
    \textbf{Sadness}& -2.50& -2.67& -1.12& -1.47& -1.73& -3.13& \textbf{-7.25}& -1.34& \textbf{-3.74}&
    \textbf{30.45}& -1.09& -0.47\\
    \textbf{Tenderness}& -2.50& -2.67& -0.14& 0.34& -1.73& -3.13& \textbf{-7.90}& \textbf{4.37}& 2.76& -3.00& -1.09& \textbf{26.18}\\
	
	\bottomrule
    \end{tabular} 
    }
    \caption{Pearson standardized residual values (Z-factor) between the target and reported discrete emotions. Values out of the range $[-3.34$, $3.34$] are in bold, meaning strong dependence.}
    \label{tab:standardizedResiduals}
\end{table}

The following conclusions are extracted from analyzing Table \ref{tab:standardizedResiduals}: (1) all the target emotions show a strong positive dependence with their corresponding reported, so the videos are especially eliciting the desired emotion in the users; (2) all videos originally classified as \textit{non-fear} have a significant negative dependence with such emotion, which implies \textit{fear} is clearly recognizable; (3) it is statistically confirmed that the clips classified as \textit{disgust, tenderness} and \textit{anger} show positive dependence with other emotions in addition to their own. Focusing on this last conclusion, it is remarkable the dependence between \textit{disgust} and \textit{amusement} since they are far from each other in the PAD space, meaning they should be easy to differentiate. For the videos classified as \textit{anger}, positive correlations are observed with the emotions of \textit{disgust} and \textit{contempt}, which fits with the behavior already obtained in previous studies \cite{preetiquetado,samClara} when it comes to clips that show gender-based violence scenes. 

The continuous PAD annotations are also examined by means of the Intra-class Correlation Coefficient (ICC) based on a single-rating, absolute-agreement, two-way mixed effects model, as implemented in the \textit{icc()} function from the library \textit{irr} in the R software. This study is used to evaluate the inter-rater consistency of the continuous annotations (targeted or reported ones) when classifying stimuli with respect to the corresponding discrete annotations (targeted or reported ones). Based on the $95\%$ confidence interval of the ICC estimate, ICC index values less than $0.5$, between $0.5$ and $0.75$, between $0.75$ and $0.9$, and greater than $0.90$ are considered poor, moderate, good, and excellent reliability, respectively \cite{ICCquality}. 

Table \ref{tab:iccPAD} shows the reliability ICC metrics analyzing the targeted continuous annotations regarding the targeted discrete ones (see the Targeted field). Analyzing this table it is found poor consistency for $4$ out of $8$ of the emotions, which are \textit{amusement, anger, disgust,} and \textit{sadness}. It highlights the case of \textit{disgust} with an even almost zero reliability. However, a moderated reliability (close to good) is obtained for \textit{fear} and \textit{tenderness}, getting good reliability for \textit{calm}. This table also shows the reliability ICC metrics analyzing the reported continuous annotations regarding the reported discrete ones (see the Reported field). Analyzing this table, it is found that the reliability obtained is slightly better than in the targeted case with poor consistency for $5$ out of $12$ of the emotions (\textit{amusement, attraction, contempt, disgust}, and \textit{sadness}), a moderated reliability for \textit{anger, hope, joy, tedium,} and \textit{tenderness}, and good reliability for \textit{calm} and \textit{fear}. From this study, we conclude that (1) the continuous labeling procedure is less robust than the discrete one (it could be due to the difficulty in understanding and applying the PAD metrics); (2) the continuous labeling for the reported emotions is slightly robust than the targeted one; and (3) fear is one of the most robust labeled continuous emotions.

\begin{table}[h!]
   \centering
   \resizebox{17cm}{!}{
   \begin{tabular}{c c c c c c c c c c c c c c}
   \toprule
	   & & \textbf{Amusement}& \textbf{Anger}& \textbf{Attraction}& \textbf{Calm}& \textbf{Contempt}& \textbf{Disgust}& \textbf{Fear}& 
	   \textbf{Hope}& \textbf{Joy}& \textbf{Sadness}& \textbf{Tedium}& \textbf{Tenderness} \\ 
    \midrule
    
    \multirow{2}{*}{\textbf{ICC index}}& \textbf{Targeted} & 0.48& 0.22& -& \textbf{0.84}& -& 0.06& 0.70& -& 0.52& 0.30& -& 0.72\\
	&\textbf{Reported}& 0.31& 0.51& 0.40& \textbf{0.85}& 0.45& 0.26& \textbf{0.81}& 0.67& 0.56& 0.38& 0.66& 0.69\\
    
	\bottomrule
    \end{tabular} 
    }
    \caption{Results of ICC calculation of continuous PAD annotations using single-rating, absolute-agreement, two-way mixed-random effects model with respect to targeted and reported discrete emotions.}
    \label{tab:iccPAD}
\end{table}


\section*{Usage Notes}


\subsection*{Emotion Recognition}

The most straightforward use of this database is the classification of the emotions felt by a woman, using machine and deep learning algorithms with unimodal and multimodal approaches.
First, the physiological signals can be used together or separately to analyze their relationship with the annotated discrete or dimensional emotions. 
The same can be done with the audio signal because even though it can be considered as part of the annotation procedure, it is not acknowledged as such, because its purpose is to have captured the last traces of the emotion elicited in the person.
Moreover, the physiological signals were recorded during the entire experiment, so that synchronization can be made with the physiological and audio signals, leading to a multi-modal or fusion scheme. On this basis, a series of experiments carried out in mono- and multi-modal emotion recognition can be found in "Supplementary Material".

The libraries recommended for further processing of the WEMAC dataset are the ones we found most useful for data cleaning and filtering for physiological and speech signals. On the one hand, Matlab\textregistered\ was employed for the physiological data processing using the TEAP toolbox \footnote{https://github.com/Gijom/TEAP}. On the other hand, for the speech signals, the librosa library\footnote{https://librosa.org/doc/latest/index.html} facilitates the loading of the speech signals and its processing, and for the extraction of the features the openSMILE library\footnote{https://audeering.github.io/opensmile-python/}, DeepSpectrum module\footnote{https://github.com/DeepSpectrum/DeepSpectrum} and the VGGish module\footnote{https://github.com/tensorflow/models/tree/master/research/audioset/vggish/} are used.

\subsection*{Accessing data - End User License Agreement (EULA)}

The use of the WEMAC dataset is licensed under a Creative Commons Attribution 4.0 International License (CC-BY-4.0). The data is hosted encrypted in the UC3M4Safety Repository in the '\textit{Consorcio Madroño}' online platform at\url{https://edatos.consorciomadrono.es/dataverse/empatia}, which includes one folder per dataset described in Table \ref{tab:database_hierarchy}. Instructions to decrypt the data will be provided after fulfilling the EULA form located on \url{https://www.uc3m.es/institute-gender-studies/DATASETS}, which should be signed and emailed to the UC3M4Safety Team (uc3m4safety@uc3m.es). 


\section*{Code availability}

The code used for the data processing and cleaning is publicly available at \url{https://github.com/BINDI-UC3M/wemac_dataset_signal_processing}. The repository contains two branches, one for the physiological signals (developed in MATLAB 2020) and one for the speech signals (developed in Python 3.6.5). All required packages are listed in the repository, which is expected to serve as a starting point for further data analyses. 

\bibliography{bibliography}

\section*{Acknowledgements} 

This work has been supported by the Dept. of Research and Innovation of Madrid Regional Authority, in the EMPATIA-CM research project (reference Y2018/TCS-5046), SAPIENTAE4Bindi Project from Grant PDC2021-121071-I00 funded by MCIN\/AEI\/10.13039/501100011033 and by the European Union ''NextGenerationEU/PRTR'',  grant PID2021-125780NB-I00 funded by AEI, and the Spanish Ministry of Science, Innovation and Universities with the FPU grant FPU19/00448. 
The authors thank all the members of UC3M4Safety for their contribution and support.

\section*{Author contributions statement}

Conceptualization and Design of the Database: All authors.
Database Capture Protocol: All authors.
Participants Assistance: C.L.O., E.R.G., E.R.P., J.M.C. and L.G.M. 
Data Curation: E.R.G., E.R.P., J.M.C. and L.G.M. 
Technical Validation: E.R.P. and L.G.M. 
Supplementary Material Experimentation: E.R.G. and J.M.C.
Original Draft Writing: E.R.G., E.R.P., J.L.G., J.M.C. and L.G.M.
Supervision: C.P.M. and C.L.O.
Review \& Editing: J.L.G, C.P.M. and C.L.O. 

\section*{Competing interests} 
The authors declare no competing interests.

\section*{Additional Information}
\textbf{Supplementary information.} The online version contains supplementary material dealing with the description of results from emotion recognition experiments. 
\textbf{Correspondence} and requests for materials should be addressed to J.M.C. and E.R.G.







\end{document}


 \section*{Supplementary material}

\newcommand{\BindiUnoCero}{Bindi 1.0}
\newcommand{\BindiDosCeroWprealarm}{Bindi 2.0a} 
\newcommand{\BindiDosCeroWOprealarm}{Bindi 2.0b} 

This supplementary material presents the baseline results for \textit{fear} detection based on the intelligence engine proposed in \cite{iot_bbddlab} by using the physiological and speech data for $87$ volunteers in WEMAC. In this baseline, $13$ of the volunteers are discarded due to high unbalance of the fear labels' distribution on them, by following a similar criterion as in \cite{iot_bbddlab}. Note that the fear detection results in \cite{iot_bbddlab} comprise the use of only $42$ volunteers of the WEMAC database.

\subsection*{Detailed Results of Fear Classification}

\begin{figure*}[h]
  \begin{subfigure}[b]{.5\textwidth}
    \includegraphics[width=\textwidth]{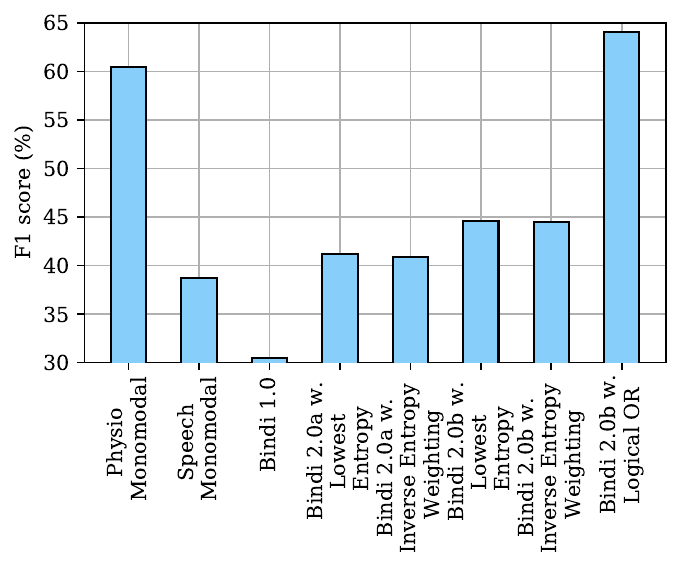}
    \caption{}
    \label{fig:f1_50}
  \end{subfigure}
  \hfill
  \begin{subfigure}[b]{.5\textwidth}
    \includegraphics[width=\textwidth]{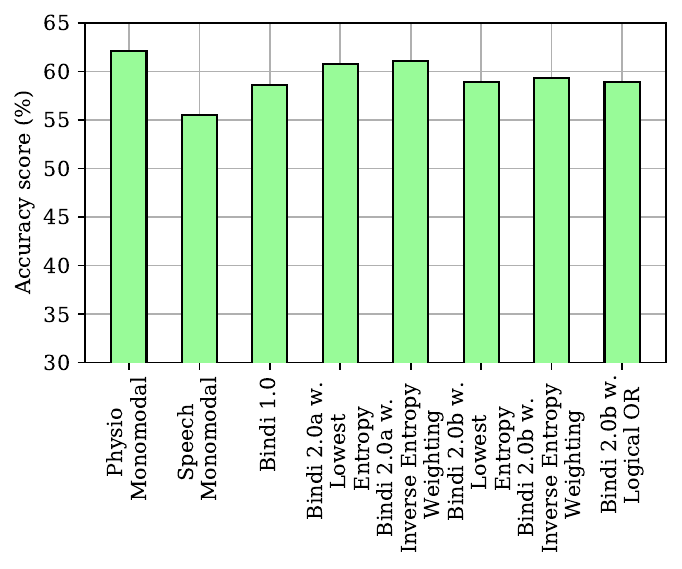}
    \caption{}
    \label{fig:acc_50}
  \end{subfigure}
  \caption{Average performance analysis predicting over the $87$ testing groups for the different architecture configurations: a) F1 score, b) Accuracy score. {From left to right, the configurations are: physiological monomodal subsystem, the speech monomodal subsystem,} \BindiUnoCero, \BindiDosCeroWprealarm\ {with lowest entropy data fusion,} \BindiDosCeroWprealarm\ {with inverse entropy weighting data fusion,} \BindiDosCeroWOprealarm\ {with lowest entropy data fusion,} \BindiDosCeroWOprealarm\ {with inverse entropy weighting data fusion, and} \BindiDosCeroWOprealarm\ {with logical OR data fusion. Note that} \BindiDosCeroWprealarm\ {was not combined with logical OR data fusion because it is equivalent to} \BindiUnoCero.}
    \label{fig:results_mono_and_fusion}
\end{figure*}

Based on the same mono-modal and data fusion architectures presented in \cite{iot_bbddlab}, the same time arrangements used for the alignment of the physiological and speech signals (\BindiUnoCero, \BindiDosCeroWprealarm, \BindiDosCeroWOprealarm) are employed in this case. Figure \ref{fig:results_mono_and_fusion} represents the F1-score (\ref{fig:f1_50}) and Accuracy (\ref{fig:acc_50}) results by considering the $87$ volunteers in WEMAC for the mono-modal fear detection systems (only based on physiological or speech data) in addition to the fusion strategies for the merging of both modalities. Both performance metrics are given to get more interpretable results from the slight imbalance between the binary \textit{fear} labels.

\begin{table*}[h]
\centering
\def\arraystretch{1.2}
\resizebox{\textwidth}{!}{%
\begin{tabular}{cc|c|c|c|c|c|c|c|c|}
\cline{3-10}
&                                                         & \textbf{\begin{tabular}[c]{@{}c@{}}Physiological\\ Monomodal\end{tabular}} & \textbf{\begin{tabular}[c]{@{}c@{}}Speech\\ Monomodal\end{tabular}} & \textbf{\begin{tabular}[c]{@{}c@{}}BINDI\\ 1.0\end{tabular}} & \textbf{\begin{tabular}[c]{@{}c@{}}\BindiDosCeroWprealarm\\\ Lowest\\ Entropy\end{tabular}} & \textbf{\begin{tabular}[c]{@{}c@{}}\BindiDosCeroWprealarm\\\ Inverse \\Entropy \\Weighting \end{tabular}} & \textbf{\begin{tabular}[c]{@{}c@{}}\BindiDosCeroWOprealarm\\\
Lowest\\ Entropy\end{tabular}} & \textbf{\begin{tabular}[c]{@{}c@{}}\BindiDosCeroWOprealarm\\\
Inverse \\Entropy\\ Weighting\end{tabular}} & \textbf{\begin{tabular}[c]{@{}c@{}}\BindiDosCeroWOprealarm\\\ Logical\\ OR\end{tabular}} \\ \hline
\multicolumn{1}{|c|}{\multirow{3}{*}{\textbf{F1-score}}} & Mean                                                    & 60.45                                                                      & 38.75                                                               & 30.46                                                        & 41.17                                                                        & 40.91                                                                          & 44.56                                                                      & 44.46                                                                        & 64.06                                                                 \\ \cline{2-10} 
\multicolumn{1}{|c|}{}                                   & \begin{tabular}[c]{@{}c@{}} \footnotesize{Std}\end{tabular}  & 17.70                                                                      & 29.26                                                               & 30.06                                                        & 29.47                                                                      & 29.60                                                                          & 29.95                                                                      & 30.01                                                                       & 14.94                                                                 \\ \hline
\multicolumn{1}{|c|}{\multirow{3}{*}{\textbf{Accuracy}}} & Mean                                                    & 62.07                                                                      & 55.50                                                                & 58.62                                                        & 60.76                                                                      & 61.08                                                                          & 58.95                                                                      & 59.28                                                                        & 58.95                                                                   \\ \cline{2-10} 
\multicolumn{1}{|c|}{}                                   & \begin{tabular}[c]{@{}c@{}} \footnotesize{Std} \end{tabular}  & 16.04                                                                      & 15.98                                                               & 14.04                                                       & 14.35                                                                       & 15.11                                                                          & 17.17                                                                      & 17.13                                                                       & 16.32                                                                 \\ \hline
\end{tabular}}
\caption{Average performance analysis predicting over the 88 testing groups. Mean and standard deviations (Std).}
\label{table:mean_and_std_results}
\end{table*}

Compared to the results in \cite{iot_bbddlab}, we now double the amount of user data, adding $45$ volunteers in these experiments. We use a \textit{LASO} (Leave hAlf Subject Out) approach in which we train $87$ models, one per each group or volunteers, using as fine-tuning data half of the data belonging to each user, and using as blind testing data the other half. Note that this is done with the intention of developing a general \textit{fear} detection model personalized to each particular user.

As for the performances, these results are very similar to the ones achieved in \cite{iot_bbddlab}, even slightly lower in some cases. This leads us to think that the addition of more data by doubling the number of users is still far from achieving higher and more reliable rates for the detection of \textit{fear}, leaving the door open for the research community to test new fusion methods, personalisation strategies for each user, study the correlation of temporal alignment of the two data modalities available, and test other suitable methodologies for this particular dataset. 

 \bibliography{bibliography}